\documentclass[11pt]{article}
\usepackage{algorithm}
\usepackage{algpseudocode}
\usepackage{amsmath}
\usepackage{amssymb}
\usepackage{amsthm}
\usepackage{arydshln}
\usepackage{authblk}
\usepackage{bm}
\usepackage{booktabs}
\usepackage[margin=1in]{geometry}
\usepackage{graphicx}
\usepackage{makecell}
\usepackage{multirow}
\usepackage{pdflscape}
\usepackage{tabularx}
\usepackage{xcolor}
\usepackage[colorlinks=true,citecolor=blue]{hyperref}

\usepackage{amsopn}

\graphicspath{{figs/}}

\title{Neural-operator element method: Efficient and scalable finite element method enabled by reusable neural operators}
\author[1,2]{Weihang Ouyang}
\author[3]{Yeonjong Shin}
\author[1]{Si-Wei Liu}
\author[2,*]{Lu Lu}
\affil[1]{Department of Civil and Environmental Engineering, Hong Kong Polytechnic University, Hung Hom, Kowloon, Hong Kong, China}
\affil[2]{Department of Statistics and Data Science, Yale University, New Haven, CT, USA}
\affil[3]{Department of Mathematics, North Carolina State University, Raleigh, NC, USA}
\affil[*]{Corresponding author. Email: lu.lu@yale.edu}

\date{}

\begin{document}
\maketitle

\begin{abstract}
The finite element method (FEM) is a well-established numerical method for solving partial differential equations (PDEs). However, its mesh-based nature gives rise to substantial computational costs, especially for complex multiscale simulations. Emerging machine learning-based methods (e.g., neural operators) provide data-driven solutions to PDEs, yet they present challenges, including high training cost and low model reusability. Here, we propose the neural-operator element method (NOEM) by synergistically combining FEM with operator learning to address these challenges. NOEM leverages neural operators (NOs) to simulate subdomains where a large number of finite elements would be required if FEM was used. In each subdomain, an NO is used to build a single element, namely a neural-operator element (NOE).
NOEs are then integrated with standard finite elements to represent the entire solution through the variational framework.
Thereby, NOEM does not necessitate dense meshing and offers efficient simulations.
We demonstrate the accuracy, efficiency, and scalability of NOEM by performing extensive and systematic numerical experiments, including nonlinear PDEs, multiscale problems, PDEs on complex geometries, and discontinuous coefficient fields.
\end{abstract}


\section{Introduction}
\label{sec:intro}

Efficient and scalable numerical simulation methods are crucial for both scientific research and engineering practice. Extensive efforts since the last century have led to the development of various simulation techniques, which can broadly be categorized into two groups: (1) mesh-based methods \cite{eymard2000fvm, bathe2006finite, leveque2007fdm}, such as the finite element method (FEM), and (2) mesh-free methods \cite{monaghan2012smoothed, yreux2017quasi}, including those based on machine learning (ML)~\cite{zhang2019quantifying,pang2019fpinns,lu2021deepxde,yu2022gradient,karniadakis2021physics,wu2023comprehensive}. Despite these advancements, achieving accuracy, efficiency, and scalability in large-scale simulations of complex multiscale systems remains significantly challenging.

\begin{figure}[htbp]
    \label{fig:method_ill}
    \centering
    \includegraphics[width=0.9\textwidth]{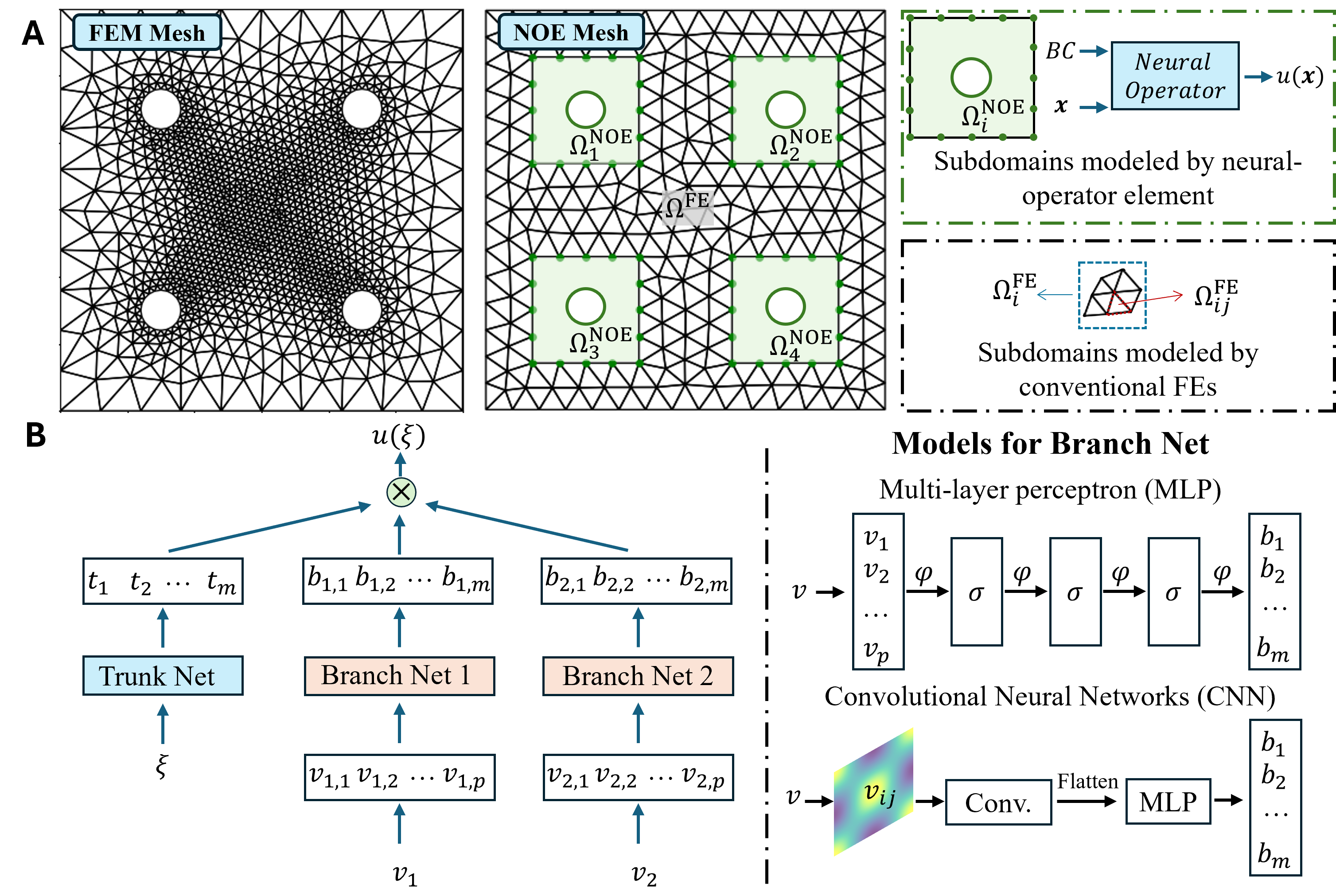}
    \caption{\textbf{Illustration of NOEM.} (\textbf{A}) Comparison between FEM and NOEM. (\textbf{B}) Illustration of the neural operator models via the multiple-input deep operator network (MIONet) \cite{jin2022mionet}. MIONet simplifies to the deep operator network (DeepONet) \cite{lu2021learning} when only one branch network is employed. Different models can be used for the branch network, such as multi-layer perceptrons (MLP) and convolutional neural networks (CNN).}
\end{figure}

In many science and engineering disciplines, FEM is a popular numerical method for solving partial differential equations (PDEs), due to its versatility and robustness \cite{belardi2021analysis, liu2020improved, naser2021modeling, li2007finite}.
However, when it comes to systems exhibiting, for example, multiscale features or defined on complex geometries (e.g., non-polygon shapes), FEM requires a large number of elements (mesh) 
(Fig.~\ref{fig:method_ill}A) to produce accurate solutions \cite{bathe2006finite, huang2021convergence}. 
Such a mesh-dependent nature makes FEM computationally expensive and limits its applicability for large-scale problems. 
While considerable efforts have been made to address this challenge, most involve a trade-off between computational efficiency and accuracy \cite{lucia2004reduced, li2021reduced}, or are only applicable for a special class of problems \cite{lee2009sem, hafeez2023sem, ouyang2024refined, scott2006plastic, ko2017shell, zou2021shell}.

On the other hand, operator learning has emerged as a new ML framework that aims to learn mappings between infinite-dimensional function spaces \cite{chen1995universal,jiao2021one,wu2025coast,wang2025fundiff}. It differentiates itself from traditional ML tasks, such as function approximation which only learns mappings between finite-dimensional spaces.
Due to its close connections to PDE solution operators, it has garnered huge attention from the scientific computing community, and many neural network models for operator learning, namely neural operators (NOs), have been proposed. Among them, deep operator networks (DeepONets) \cite{lu2021learning}, as the first NO model, are inspired by the universal operator approximation theorem \cite{chen1995universal}.
DeepONets have demonstrated superior performance in many challenging tasks \cite{lu2022multifidelity,goswami2022physics, azizzadenesheli2024neural, he2023novel, lu2021learning, kontolati2024latenddno,lee2024training}.
In addition, the inherent structure from the universal approximation theorem offers significant flexibility in designing sub-architectures~\cite{jiang2024fourier,lee2024efficient,yin2024scalable,xiao2025quantum}. 
Despite empirical successes, operator learning faces multiple challenges for large-scale simulations \cite{niaki2021physics, yan2022framework,kou2025efficient,zhang2024federated}, such as high-fidelity data generation, demanding training costs, and reusability. 
To overcome the limitations of the pure data-driven nature, physics-informed ML (PIML) has the potential to circumvent the need for a large amount of data by incorporating PDEs into the loss \cite{jagtap2020conservative, karniadakis2021physics, lu2021deepxde,jiao2024solving}. However, physics-informed training remains difficult as the loss involved with PDEs often incurs new challenges \cite{jagtap2020extended, kharazmi2021hp, ainsworth2022active}, especially for large-scale complex systems.

In the present work, we propose a hybrid method, namely, the neural-operator element method (NOEM; Fig.~\ref{fig:method_ill}A). The motivation is to combine the strengths of FEM and operator learning to mitigate their limitations. The underlying idea of NOEM is to represent the PDE solution using both finite elements and neural-operator elements (NOEs). NOEs are constructed by pretrained NOs and assigned to parts of computational domains where a large number of finite elements would be required if FEM was used.
Each NOE is assigned to a specific non-overlapping subdomain, and NOEM solves the energy norm minimization problem, which is similar to FEM. The adoption of NOEs substantially reduces the complexity of the problem, enabling NOEM to be applied effectively to large-scale problems. Extensive numerical examples are performed to validate the proposed NOEM and illustrate the key features of the NOEM (Table~\ref{table:ex_res_sum}). 

\setlength{\extrarowheight}{1pt}
\begin{table}[htbp]
\centering
\caption{\textbf{Summary of key features and results.}}
\label{table:ex_res_sum}
\footnotesize
\begin{tabular}{ccl}
\toprule
\multicolumn{2}{c}{\textbf{Features}}                                                            & \textbf{Description \& Key results}                                                         \\ \midrule
\multirow{4}{*}{\begin{tabular}{@{}c@{}} \textit{\textbf{Compared}} \\  \textit{\textbf{to FEM}}\end{tabular}} 
                                                   & \multirow{2}{*}{\textbf{Efficiency}}        & \textbf{\textit{Reduce computational time}}                        \\
                                                   &                                             & $\bullet$ Section \ref{sec:ex3.2}: Boosted computational speed by around 14 times over FEM                              \\ \cline{2-3} 
                                                   & \multirow{2}{*}{\textbf{Scalability}}       & \textbf{\textit{Support effective large-scale domain modeling}}                                                      \\
                                                   &                                             & $\bullet$ Section \ref{sec:ex3}: Analysis across varying domain sizes, scaling up by 100 times                   \\ \midrule
\multirow{4}{*}{\begin{tabular}{@{}c@{}} \textit{\textbf{Compared}} \\  \textit{\textbf{to NO}}\end{tabular}}  
                                                   & \multirow{2}{*}{\begin{tabular}{@{}c@{}} \textbf{Training} \\  \textbf{Efficiency}\end{tabular}} 
                                                   & \multirow{2}{*}{\begin{tabular}{@{}l@{}} \textbf{\textit{Require smaller training datasets and faster training processes than}} \\  \textbf{\textit{direct modeling of entire systems using NOs}}\end{tabular}}\\  
                                                   &                                             &  \\ \cline{2-3} 
                                                   & \multirow{3}{*}{\begin{tabular}{@{}c@{}} \textbf{Model} \\  \textbf{Reusability}\end{tabular}} 
                                                                                                 & \textbf{\textit{Directly reuse pretrained NOs across varied problems}}                  \\
                                                   &                                             & $\bullet$ Section \ref{sec:ex1}: Single NO for 2000 problems of varying coefficient functions               \\
                                                   &                                             & $\bullet$ Section \ref{sec:ex3}: Single NO for 200 problems with varying geometries                     \\  \midrule
\multirow{5}{*}{\textit{\textbf{Others}}}          & \multirow{2}{*}{\begin{tabular}{@{}c@{}} \textbf{ML} \\  \textbf{Compatibility}\end{tabular}}     & \textbf{\textit{Achieve compatibility with different NO models}}                                   \\
                                                   &                                             & $\bullet$ Section \ref{sec:exs}: Various utilized NOs, such as DeepONet, MIONet, and CNN \\ \cline{2-3} 
                                                   & \multirow{3}{*}{\begin{tabular}{@{}c@{}} \textbf{PDE} \\  \textbf{Applicability}\end{tabular}}     & \textbf{\textit{Enable application to various PDE problems}}                                          \\
                                                   &                                             & $\bullet$ Section \ref{sec:ex2}: Multiscale problems                                                            \\
                                                   &                                             & $\bullet$ Section \ref{sec:ex4.4}: Nonlinear problems                                                               \\ \bottomrule
                                                   
\end{tabular}
\end{table}

Hybridizing FEM and ML is not new and has been explored in the literature. 
For example, Refs.~\cite{saha2021hidnn, wang2023epinn, rezaei2024fol, zang2020weak} use piecewise polynomials and weak-form residuals in designing ML architectures and physics-informed loss functions, which are found to be effective to some extent. However, the reusability of these models remains elusive, especially for large-scale systems.
Ref.~\cite{yin2022dd} has integrated NOs into the domain decomposition (DD) approach of FEM by modeling different regions separately with NOs and finite elements. Ref.~\cite{chung2024dgdd} also embedded learned reduced-order models (ROM) within the discontinuous Galerkin domain decomposition (DG-DD) framework, enabling efficient simulation by applying ROMs to unit components and interfaces. Despite these advances, modeling large-scale systems with multiple subdomains using DD is still challenging, as it commonly requires tedious iteration processes and may fail to converge if certain parameters are not carefully chosen \cite{yin2022dd, ernst2011difficult}. 

What sets NOEM asides from existing hybrid methods are (1) the approach of learning NO elements that cover a large subdomain and (2) the use of standard FEs for subdomains that FEM works well. Particularly, this feature is preferred in industrial applications. The contemporary industry practice frequently employs a ``standardized product module'' philosophy, constructing large-scale systems by interconnecting standardized modules or components, e.g., modular integrated construction \cite{wuni2020mic}. In this context, NOEM offers pre-trained NO libraries, i.e., NOEs, tailored for simulating standardized modular components across various projects.

Additionally, the inherent differentiability of NOEs provides differentiable modeling approach for large-scale systems. While not the focus of this paper, such differentiable feature holds significant potential for optimal design tasks.
\section{Preliminaries}

In this section, we first briefly revisit several NOs, including DeepONet and its variants, and then review FEM, covering both the standard weak form and the variational form. Finally, we discuss the enforcement of hard constraints on NOs under varying boundary conditions.

\subsection{Neural operators}
\label{sec:nom}

Let $\mathcal{G}:V \to H$ be an operator of interest where $V$ and $H$ are the spaces of the input and output functions, respectively. Each function in $V$ and $H$ is real-valued defined on $\Omega' \subset \mathbb{R}^{d'}$ and $\Omega \subset \mathbb{R}^{d}$, respectively. The goal of operator learning is to construct a NO, $\mathcal{G}_\text{NO}$, from a set of training data consisting of pairs $(v_j, u_j =\mathcal{G}[v_j])$ in discretized forms. For the sake of simplicity, we slightly abuse notation and refer to a discretized form of $v$ as $v$.
We revisit three DeepONet variants: the vanilla DeepONet \cite{lu2021learning}, the convolutional DeepONet \cite{lu2022comprehensive, mei2024fully}, and the multi-input DeepONet (MIONet) \cite{jin2022mionet}.
These NOs are introduced and employed in the present study given their superior generalization capability \cite{goswami2022physics, azizzadenesheli2024neural, he2023novel}.

\paragraph{Vanilla DeepONet.}
DeepONet \cite{lu2021learning} constructs $\mathcal{G}_\text{NO}$ through two distinct sub-networks: a branch and a trunk network (Fig.~\ref{fig:method_ill}B).
The branch network takes a discretized function $v$ as input and outputs a vector $\bm{b}(v)=[b_0, b_1(v), \dots, b_m(v)]^\top$, and the trunk network takes a point $x \in \Omega$ as input and outputs a vector $\bm{t}(x)=[1, t_1(x), \dots, t_m(x)]^\top$.
DeepONet is then represented by the Euclidean inner product of these outputs:
\[
\mathcal{G}_\text{NO}[v](x) = \langle \bm{b}(v), \bm{t}(x)\rangle = \sum_{i=1}^m b_i(v) t_i(x) + b_0.
\]
MLPs are often used as the default for both subnetworks. If this is the case, we refer to DeepONet as vanilla DeepONet.

\paragraph{Convolutional DeepONet.}
Convolutional DeepONet \cite{lu2022comprehensive, mei2024fully} is a variant of vanilla DeepONet, which uses a convolutional neural network (CNN) as a branch network (Fig.~\ref{fig:method_ill}B). This variant is found to be particularly effective when the discretized input function has a substantially large size and has an image-like shape.

\paragraph{Multi-input DeepONet.}
MIONet \cite{jin2022mionet} is designed to learn an operator that takes several input functions. Similar to DeepONet, the architecture of MIONet is based on the universal operator approximation theorem, which extends from single to multiple Banach spaces. The operator of interest is $\mathcal{G}:V_1\times \cdots \times V_n \to H$, where $V_1,\dots,V_n$ are $n$ input function spaces and each $v_i \in V_i$ is a real-valued function defined on $\Omega_i' \subset \mathbb{R}^{d_i'}$. To handle multiple inputs, MIONet employs multiple branch networks, each corresponding to a distinct input function space, and processes them independently while keeping a single trunk network (Fig.~\ref{fig:method_ill}B). MIONet is represented as follows:
\begin{equation*} 
\mathcal{G}_\text{NO}[v_1, v_2, \dots, v_n](x) = \sum_{i=1}^m \left( \prod_{j=1}^n b_{i,j}(v_j) \right) t_i(x) + b_0,
\end{equation*} 
where $b_{i,j}$ represents the $i$\textsuperscript{th} output feature from the $j$\textsuperscript{th} branch network.

\subsection{Neural operators with hard-constraint boundary conditions}

Let us consider a general PDE defined in $\Omega$ with a Dirichlet boundary condition:
\begin{equation}\label{eq:hc_pde_setting}
	u = g       \quad \text{on }   \partial \Omega.
\end{equation}

For the simplicity of discussion, let the PDE solution operator from Dirichlet boundary conditions be $\mathcal{G}: g \mapsto u$. It is often desirable to construct NOs that exactly satisfy the imposed boundary conditions. We refer to such NOs as hard-constraint NOs, which are proven to be effective in some problems \cite{lu2022comprehensive, brecht2023improving}.

Let $\alpha$ be an auxiliary function satisfying $\alpha(x) = 0$ if $x \in \partial \Omega$ and $\alpha(x)>0$ if $x \not\in \partial \Omega$, and let $\beta_g$ be an extension of $g$, i.e., $\beta_g(x) = g(x)$ if $x \in \partial \Omega$. Then hard-constraint NOs can be constructed by
\begin{equation}
    \label{eq:hcno}
    \mathcal{G}^\text{hc}_\text{NO}[g](x) = \alpha(x) \mathcal{G}_\text{NO}[g](x) + \beta_g(x) \quad \text{for } x \in \Omega,
\end{equation}
where $\mathcal{G}_\text{NO}[g](x)$ represents a NO model with learnable parameters. 
However, the construction of $\alpha$ and $\beta_g$ could be challenging.
We therefore consider an approximation approach where $\beta_g$ is approximated by a mesh-based polynomial $p_g$ constructed from a set of discretized boundary data $\{x_i \in \partial\Omega, g(x_i)\}_{i=1}^m$.
Appendix~\ref{sec:app_hcno} presents the details on constructing $\alpha$ and $\beta_g$.

\subsection{Finite element method}

This section briefly reviews FEM. Suppose that the PDE has a variational form:
\begin{equation}\label{eq:weak}
	\text{Find } u\in H, \text{ such that } 
	\quad a(u,v) = L(v) \quad\forall\,v\in H_0,
\end{equation}
where $a(\cdot,\cdot)$ and $L(\cdot)$ are the bilinear form and the linear functional, respectively, and $H_0$ is a test function space. The Lax–Milgram lemma guarantees the existence and uniqueness of the solution to Eq.~\eqref{eq:weak} under certain assumptions on $a$ and $L$. Furthermore, we assume that Eq.~\eqref{eq:weak} is equivalent to the energy minimization problem:
\begin{equation}\label{eq:energy}
	u  = \arg\min_{w\in H} J[w]:=\frac12\,a(w,w)-L(w).
\end{equation}

We consider $\{\Omega_{i}^\text{FE} \}_i$ as a partition of $\Omega$. 
Furthermore, we let $\{\Omega_{ij}^\text{FE}\}_j$ be a partition of $\Omega_i^\text{FE}$, where $\Omega_{ij}^\text{FE}$ is commonly selected as the polygon with regular geometry. FEM represents an approximation to the solution on $\Omega_i^\text{FE}$ as a linear combination of shape functions $\{p_{ij}^\text{FE}\}_j$: 
\begin{equation} \label{eq:fem}
    \phi_i^\text{FE}(x;\bm{\alpha}_i) = \sum_{j} \alpha_{ij} p_{ij}^\text{FE}(x),
\end{equation}
where $\bm{\alpha}_i=\{ \alpha_{ij}\}_j$ is the collection of $\alpha_{ij}$; and each function $p_{ij}^\text{FE}$ is supported on $\Omega_{ij}^\text{FE}$, i.e., vanishing outside of $\Omega_{ij}^\text{FE}$. In our numerical experiments, $p_{ij}^\text{FE}$ is selected as a piecewise linear function.
Then, the FEM solution defined on the entire computational domain $\Omega$ can be represented via
\begin{equation*}
    q(x;\{\bm{\alpha}_{i}\}_i) = \sum_{i} \phi_i^\text{FE}(x;\bm{\alpha}_i),
\end{equation*}
where $\{\bm{\alpha}_{i}\}_i$  is obtained by solving Eq.~\eqref{eq:energy}:
\begin{equation*}
	\{\bm{\alpha}_{i}\}_i^*  = \arg\min_{\{\bm{\alpha}_{i}\}_i} J[q(\cdot;\{\bm{\alpha}_{i}\}_i)].
\end{equation*}

\section{Methods}
\label{sec:methods}

We consider the following PDE defined on a domain $\Omega$:
\begin{equation*} \label{eq:pde}
    \mathcal{D}[u](x) = 0, \quad \text{ for } x \in \Omega,
\end{equation*}
where $\mathcal{D}$ is the underlying differential operator of the PDE. In NOEM, we partition $\Omega$ into two non-overlapping subdomains $\Omega_\text{FE}$ and $\Omega_\text{NOE}$, i.e.,
\begin{equation*}
    \Omega = \Omega^\text{FE} \cup \Omega^\text{NOE},
\end{equation*}
where $\Omega^\text{FE}=\{ \Omega^\text{FE}_i \}_{i\in I_\text{FE}}$ is a subdomain represented by classical FE meshes, and $\Omega^\text{NOE}= \{ \Omega^\text{NOE}_j \}_{j\in I_\text{NOE}}$ is represented by NOE. 
For instance, in Fig.~\ref{fig:method_ill}A, the FE subdomain $\Omega^\text{FE}$ is represented by the black grid. 
The subdomain $\Omega^\text{NOE}$ modeled by the NOEs (visualized by the green areas) is further partitioned into four subdomains as $\Omega^\text{NOE} = \{ \Omega^\text{NOE}_j \}_{j\in \{1,2,3,4\}}$, where each $\Omega^\text{NOE}_j$ is modeled by one single NOE. 
In the following content, we first introduce how to construct NOEs by NOs. Next, we present the NOEM framework for solving PDEs. Finally, a discussion of the NOEM is provided.

\subsection{NOE represented by a neural operator}
\label{sec:no_def}

Our NOEM framework is generic and can be applied to different types of PDEs. Here, to simplify the notations and explanation, we consider the case where the Dirichlet boundary condition is sufficient to guarantee the existence and uniqueness of the PDE solution. We consider the PDE restricted to a subdomain modeled by the NOE, $\Omega^\text{NOE}_j$, and then the solution inside $\Omega^\text{NOE}_j$ denoted by $u|_{{\Omega^\text{NOE}_j}}$ is determined by the boundary condition at $\partial \Omega^\text{NOE}_j$ denoted by $u|_{\partial{\Omega^\text{NOE}_j}}$. We denote this Dirichlet-to-solution operator as $\mathcal{G}: u |_{\partial \Omega^\text{NOE}_j} \mapsto u|_{\Omega^\text{NOE}_j}$.

We use a neural operator to learn the operator $\mathcal{G}$. Specifically, the Dirichlet boundary condition for the NO input is parametrized by the sensor values, $\bm{\beta}_j$, whose locations are aligned with the nodes determined by the adjacent finite elements, as illustrated by the green dots in Fig.~\ref{fig:method_ill}. The details about learning NO are presented in Appendix~\ref{sec:no_training}. We note that as $\mathcal{G}$ is the solution operator of a small subdomain, learning $\mathcal{G}$ is significantly easier compared with the full PDE solution operator. The pre-trained NO is then fixed and used in NOEM for solving a new PDE.

\subsection{Neural-operator element method}

The principle of NOEM is to replace a large number of finite elements with a single NOE. NOEM represents the solution as
\begin{equation} \label{eqn:solution}
    q(x;\bm{c}) = \sum_{i \in I_\text{FE}} \phi_i^\text{FE}(x;\bm{\alpha}_i) + \sum_{j \in I_\text{NOE}} \phi_j^\text{NOE}(x;\bm{\beta}_j),
\end{equation}
where $\bm{c} = \{\bm{\alpha}_i\} \cup \{\bm{\beta}_j\}$. $\phi_i^\text{FE}(x;\bm{\alpha}_i)$ is the standard FE representation in Eq.~\eqref{eq:fem} whose support is $\Omega^\text{FE}_i$. $\phi_j^\text{NOE}(x;\bm{\beta}_j)$ of the support $\Omega^\text{NOE}_j$ is the NOE:
\begin{equation*}
    \phi_j^\text{NOE}(\cdot;\bm{\beta}_j) = \mathcal{G}[u|_{\partial \Omega^\text{NOE}_j} ](\cdot),
\end{equation*}
where $\bm{\beta}_j$ is the discrete representation of $u|_{\partial \Omega^\text{NOE}_j}$ as defined in Section~\ref{sec:no_def}. A pedagogical example for constructing $q(x;\bm{c})$ is illustrated in Section~\ref{sec:ex1}.
We note that for each $\Omega^\text{NOE}_j$, if we use FEs, then a large number of elements would be needed. 
The degree of freedom (DOF) for $\bm{\alpha}'_j$ increases proportionally with the number of finite elements, thereby increasing computational complexity. 
However, this is not the case for NOEs, as they take the parametrized boundary conditions $\bm{\beta}_j$ as inputs whose cardinality is significantly smaller than that of $\bm{\alpha}'_j$.

The NOEM solution $q(\cdot;\bm{c}^*)$ is then obtained by solving Eq.~\eqref{eq:energy} in the mixed element space, i.e.,
\begin{equation}\label{eq:noe3}
	\bm{c}^*  = \arg\min_{\bm{c}} J[q(\cdot;\bm{c})],
\end{equation}
whose complexity is significantly reduced compared to the full FEM. 
Yet, the optimization problem in Eq.~\eqref{eq:noe3} is no longer convex. While it is not guaranteed to find a global minimizer, extensive numerical experiments indicate that it does not incur severe issues in this regard. Moreover, the implementation of the energy minimization in Eq.~\eqref{eq:noe3} requires numerical integration. Typically, the terms involving only FEs can be exactly computed due to the exactness of quadrature, and we only apply numerical quadrature rules or the Monte-Carlo method to approximate the terms of NOEs. In this study, we employ Newton's method to solve Eq.~\eqref{eq:noe3}, see details in Appendix~\ref{sec:noe_imp}.

\subsection{Discussion}

In the presentation above, we only consider boundary conditions as the inputs of $\mathcal{G}$, and it is straightforward to include additional input functions $f$, such as forcing/source terms and coefficient functions. In this case, $\mathcal{G}: (u|_{\partial{\Omega^\text{NOE}_j}}, f) \mapsto u|_{\Omega^\text{NOE}_j}$ is the solution operator that maps the boundary conditions and the additional function of interest $f$ to the PDE solution. Moreover, for some PDEs where the Dirichlet boundary condition is insufficient, such as the biharmonic equation, more boundary conditions can be taken as the input of NOs.

In this paper, all subdomains $\Omega^\text{NOE}_j$ have the same geometry and we only use one single pre-trained NO. The NOEM framework can be easily adapted to cases, where we use multiple NOs for different $\Omega^\text{NOE}_j$, e.g., when considering different subdomains characterized by different geometries.

In order for the NOEM solution to be consistent with the reference FEM solution, Eq.~\eqref{eqn:solution} is required to satisfy a regularity condition, such as $C^0$ or $C^1$ continuities. Therefore, in principle, the NOs with hard-constraint boundary conditions in Eq.~\eqref{eq:hcno} should be used for NOEs. However, constructing them poses another computational challenge and costs in practice. In our numerical experiments, we show that NOs with soft-constraint boundary conditions also yield satisfactory performance. This resembles the idea of the discontinuous Galerkin approach.

\section{Results}
\label{sec:exs}

In this section, ten examples in four groups are conducted to validate the performance of the proposed NOEM. For pedagogical purposes, the first example (Section~\ref{sec:ex1}) presents a simple ODE problem to illustrate the implementation of NOEM. The second example (Section~\ref{sec:ex2}) demonstrates the potential of NOEM in addressing multiscale problems and illustrates its scalability. The third example (Section~\ref{sec:ex3}) illustrates the reusability of the NO models within the NOEM. Additionally, this example also demonstrates the computational efficiency and scalability of NOEM on two-dimensional problems. Finally, the last example (Section~\ref{sec:ex4}) investigates different types of Darcy flow problems, showcasing the ability of NOEM to handle highly nonlinear problems. The codes in this study will be publicly available at the GitHub repository \url{https://github.com/lu-group/noem} once the paper is published.

\subsection{Pedagogical example}
\label{sec:ex1}

We first use a 1D elliptic ODE as a simple example to illustrate the implementation of NOEM for pedagogical purposes. The governing equation and boundary conditions are detailed as follows:
\begin{equation*}
\label{eq:ex1_setup}
\begin{aligned}
    \frac{d}{dx}\left(k(x) \frac{du}{dx}\right) = 0, &\quad x \in \Omega= [0,3], \\
    u(x) = 0,  &\quad \text{at } x = 0, \\
    \frac{du}{dx} = 0.1, &\quad \text{at } x = 3,
\end{aligned}
\end{equation*}
where
\begin{equation*}
    k(x) = \begin{cases} 
1  &\quad \text{for } x \in [0,1) \cup (2,3], \\
k'(x)  &\quad \text{for } x \in [1,2],
\end{cases}\\
\end{equation*}
and $k'(x)$ is a nonlinear coefficient function.

\begin{figure}[htbp]
    \centering
    \includegraphics[width=\linewidth]{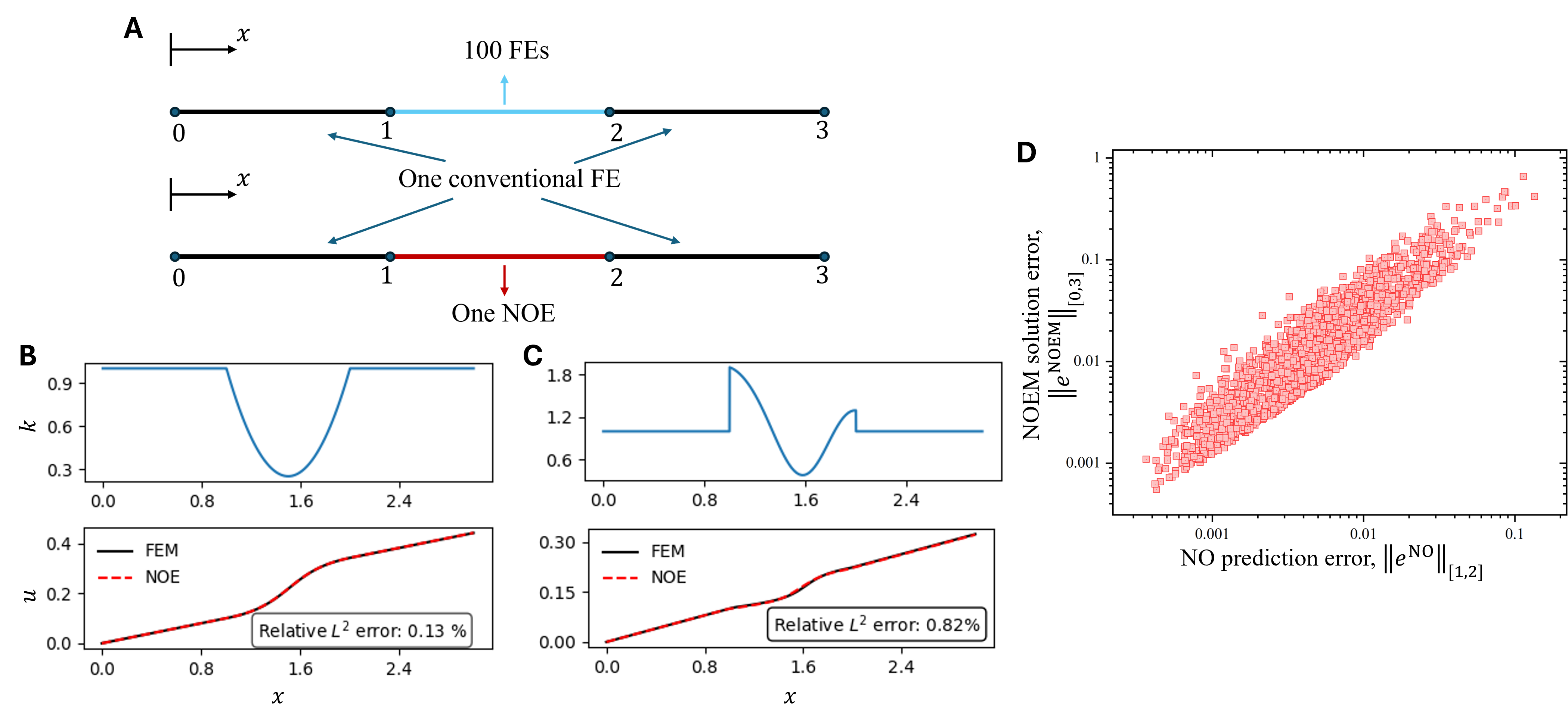}
    \caption{\textbf{Two modeling methods and numerical results in Section~\ref{sec:ex1}.} (\textbf{A}) Two modeling methods used in the conventional FEM and the NOEM are visualized in the first and second rows, respectively. The first and last segments (black lines) are modeled by one conventional 1D linear FE. The conventional FEM and the NOEM use 100 elements and one NOE to model the middle segment, respectively. (\textbf{B}) The quadratic coefficient function in Section~\ref{sec:ex1.1} and the corresponding results of the FEM and the NOEM. (\textbf{C}) The coefficient function sampled from the Gaussian process in Section~\ref{sec:ex1.2} and the corresponding results of the FEM and the NOEM. 
    (\textbf{D}) 2000 $k'(x)$ are randomly sampled. For each $k'$, we show the NO prediction error ($\|e^\text{NO}\|_{[1,2]}$) and the NOEM solution error ($\|e^\text{NOEM}\|_{[0,3]}$).
    }
    \label{fig:ex1_fig1}
\end{figure}

For the interval $[0,1)$ or $(2,3]$, we use only one conventional linear FE for each interval to accurately recover the solutions (Fig.~\ref{fig:ex1_fig1}A). For the middle segment $[1,2]$, two different modeling methods are used for comparison: (1) the conventional FEM using 100 FEs, and (2) the proposed NOEM using one NOE.

\subsubsection{A quadratic coefficient function}
\label{sec:ex1.1}

We first select $k^\prime(x)$ as a quadratic function, $k^\prime\left(x\right)=1+3(x-1)(x-2)$. To model the middle segment $[1,2]$, a DeepONet is trained to approximate the mapping from the Dirichlet boundary conditions of the following PDEs to the corresponding solutions:
\begin{equation*}
\begin{aligned}
\frac{d}{dx}\left(k'(x) \frac{du_{\mathcal{G}}}{dx}\right) &= 0, \quad &&\text{for }x \in [1,2], \\
u_{\mathcal{G}}(x)&=u_{B,1}, \quad &&\text{at }x = 1, \\
u_{\mathcal{G}}(x)&=u_{B,2}, \quad &&\text{at }x = 2,
\end{aligned}
\end{equation*}
\begin{equation*}
\mathcal{G}: \boldsymbol{u}_{\text{BC}} \mapsto u_{\mathcal{G}},
\end{equation*}
where $\boldsymbol{u}_{\text{BC}}=\{u_{B,1}, u_{B,2}\}$ is the boundary condition.

The NOEM only requires four DOFs to represent the PDE solution: $\boldsymbol{c}=\{u_i\}_{i=0}^3$, where $u_i$ is the solution value at the mesh point $x_i=i$ (Fig.~\ref{fig:ex1_fig1}A, blue dots). $\{u_0, u_1\}$ and $\{u_2, u_3\}$ represent the elements for the conventional FEs modeling the first and last segments. $\{u_1, u_2\}$ corresponds to the DOFs for the NOE, which also serve as the input $\boldsymbol{u}_{\text{BC}}$ to the branch net when formulating the NOE using the DeepONet. Specifically, according to Eq.~\eqref{eqn:solution}, the NOEM represents the solution as
\begin{equation*}
    q(x;\bm{c}) = \phi_1^\text{FE}(x;\bm{\alpha}_1) + \phi_1^\text{NOE}(x;\bm{\beta}_1) + \phi_2^\text{FE}(x;\bm{\alpha}_2) ,
\end{equation*}
where $\bm{\alpha}_1=\{u_0, u_1\}$ and $\bm{\alpha}_2 = \{u_2, u_3\}$ are the coefficients for the FEs; and $\bm{\beta}_1 = \{u_1, u_2\}$ parametrizes the NOE representation. $\phi_1^\text{FE}$, $\phi_2^\text{FE}$, and $\phi_1^\text{NOE}$ are
\begin{equation*}
    \begin{aligned}
        \phi_1^\text{FE}(x;\bm{\alpha}_1) &= 
    \begin{cases}
        (u_1 - u_0)x + u_0   &\quad \text{for } x \in \Omega_1^\text{FE}, \\
        0  &\quad \text{for } x \in \Omega \setminus \Omega_1^\text{FE},
    \end{cases} \\
    \phi_2^\text{FE}(x;\bm{\alpha}_2) &= 
    \begin{cases}
        (u_3 - u_2)(x-2) + u_2   &\quad \text{for } x \in \Omega_2^\text{FE}, \\
        0  &\quad \text{for } x \in \Omega \setminus \Omega_2^\text{FE},
    \end{cases} \\
    \phi_1^\text{NOE}(x;\bm{\beta}_1) &= 
    \begin{cases}
        \mathcal{G}[\bm{\beta}_1](x)   &\quad \text{for } x \in \Omega_1^\text{NOE}, \\
        0  &\quad \text{for } x \in \Omega \setminus \Omega_1^\text{NOE},
    \end{cases}\\
    \end{aligned}
\end{equation*}
where $\Omega_1^\text{FE}$, $\Omega_2^\text{FE}$, and $\Omega_1^\text{NOE}$ are $[0,1]$, $[2,3]$, and $[1,2]$, respectively.

The PDE solutions obtained from FEM and NOEM are in Fig.~\ref{fig:ex1_fig1}B. The NOEM achieves similar relative error (0.1\%) as FEM, even when using only one NOE to model the middle segment with the nonlinear coefficient function.

\subsubsection{Random coefficient functions sampled from a Gaussian process}
\label{sec:ex1.2}

We further test NOEM on Eq.~\eqref{eq:ex1_setup} with $k^\prime(x)$ randomly sampled from a Gaussian process (GP):
\begin{equation*}
\begin{aligned}
k^\prime(x) &= \exp(v(x)), \\
v(x) &\sim \mathcal{GP}(0, k_l(x, x^\prime)), \\
k_l(x, x^\prime) &= -\frac{1}{2}\left(\frac{x - x^\prime}{l}\right)^2,
\end{aligned}
\end{equation*}
where $\mathcal{GP}$ denotes a GP with the kernel function $k_l$, and $l=0.3$ is the correlation length. We use NOEM to solve the PDE for any $k'(x)$.

NOEM here works very similarly as the one for the previous quadratic coefficient function. The only difference is that $k'(x)$ is not fixed, and thus $k'(x)$ is also one input of $\mathcal{G}$, i.e., $\mathcal{G}: (k', \boldsymbol{u}_{\text{BC}}) \mapsto u_{\mathcal{G}}$. We use MIONet to learn this solution operator. One example of $k^\prime$ is shown in Fig.~\ref{fig:ex1_fig1}C, and FEM and NOEM have very similar PDE solutions with 0.82\% relative error.

Further tests are conducted to investigate the relationship between the error of the pre-trained NO models and final NOEM solutions. We randomly sample 2000 $k'(x)$ functions from the GP. For each $k'$, the true solution $u_{k'}$ and the NOEM solution $u_{k'}^\text{NOEM}$ are computed, and we compute the NOEM solution error as $e^\text{NOEM} = u_{k'} - u_{k'}^\text{NOEM}$. Next, using $\boldsymbol{u}_{\text{BC}} = \{u_{k'}(1), u_{k'}(2)\}$ and $k'$ as inputs to the pretrained NO, the NO prediction is $u_{k^\prime}^\text{NO}$, so the error for the NO prediction is $e^\text{NO} = u_{k'} - u_{k'}^\text{NO}$. We note that the $L^2$ norms of the errors of the NOEM and NO are evaluated over the intervals $[0,3]$ and $[1,2]$, denoted by $\|e^\text{NOEM}\|_{[0,3]}$ and $\|e^\text{NO}\|_{[1,2]}$, respectively. We find that there is a strong linear correlation between the errors of the NOEM and the NO (Fig.~\ref{fig:ex1_fig1}D).

\subsection{1D multiscale problems}
\label{sec:ex2}
In this example, the performance of the NOEM is evaluated through two 1D multiscale problems.

\subsubsection{Multiscale coefficient function}
\label{sec:ex2.1}

\begin{figure}[htbp]
    \centering
    \includegraphics[width=\linewidth]{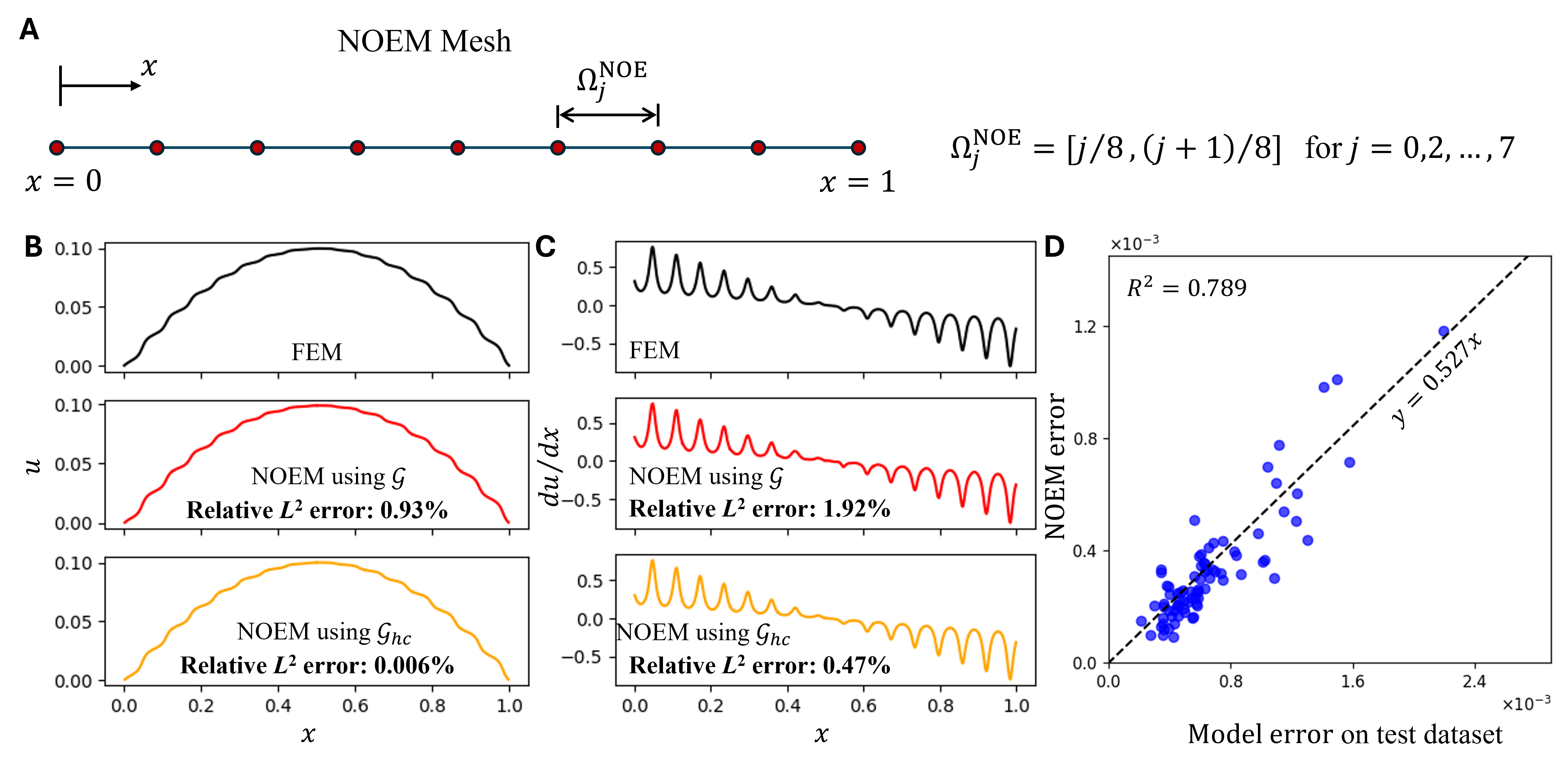}
    \caption{\textbf{Multiscale coefficient problem (Section~\ref{sec:ex2.1}).} 
    (\textbf{A}) The mesh employed for the NOEM.
    (\textbf{B}) The PDE solutions from the FEM, the NOEM using the vanilla DeepONet ($\mathcal{G}$), and the NOEM using DeepONet with the hard-constraint BC ($\mathcal{G}_{hc}$). 
    (\textbf{C}) The derivative of the PDE solutions from the three methods.
    (\textbf{D}) 80 different $\mathcal{G}_{hc}$ are trained independently with different settings. The performance of each NO on the test dataset and the performance of the NOEM when using the corresponding NO is illustrated. The $x$-axis shows NO test performance (mean relative $L^2$ error on $[0, \tfrac{1}{8}]$ with 100 boundary conditions). The $y$-axis shows corresponding NOEM performance (mean relative $L^2$ error for Eq.~\eqref{eq:ex2.1_setup} on $[0, 1]$ with 100 boundary conditions).}
    \label{fig:ex21_results}
\end{figure}

We first consider a 1D problem with multiscale coefficient functions~\cite{zhang2024bayesian}:
\begin{equation}
\label{eq:ex2.1_setup}
\begin{aligned}
    -\frac{d}{dx}\left(a\left(\frac{x}{\epsilon}\right) \frac{du}{dx}\right) &= f(x), \quad x \in [0,1], \\
a\left(\frac{x}{\epsilon}\right) = 0.5 \times &\sin\left(2\pi \frac{x}{\epsilon}\right) + 0.8, \\
u(x) = u_1,  &\quad \text{at } x = 0, \\
u(x) = u_2,  &\quad \text{at } x = 1, \\
\end{aligned}
\end{equation}
where $u_1=u_2=0$, $f(x)=0.5$, and $\epsilon = \frac{1}{16}$. In the NOEM, the domain of $[0, 1]$ is divided into 8 NOEs of length in $\frac{1}{8}$ (Fig.~\ref{fig:ex21_results}A). We train a NO to approximate the solution operator of interval subdomains with length $\frac{1}{8}$: 
\begin{equation*}
\label{eq:ex2.1_no_model}
\begin{aligned}
    \mathcal{G}: \boldsymbol{u}_{\text{BC}} \mapsto u_{\mathcal{G}},
\end{aligned}
\end{equation*}
where $\boldsymbol{u}_{\text{BC}}=\{u_{B,1}, u_{B,2}\}$ are the left and right boundary conditions of the subdomain.

We test NOEM with DeepONet trained by soft-constraint boundary conditions (denoted by $\mathcal{G}$) and with DeepONet trained by hard-constraint boundary conditions (denoted by $\mathcal{G}_{hc}$). Compared to FEM, the NOEM using different NO models both yield accurate results for this multiscale problem with respect to both the solution value (Fig.~\ref{fig:ex21_results}B) and the derivative value (Fig.~\ref{fig:ex21_results}C).

Next, we investigate the relationship between the accuracy of NO and the accuracy of the NOEM solution. We trained 80 DeepONets with hard-constrained BC of different network widths and depths from different training dataset sizes. Each NO is tested on a dataset comprising 100 samples randomly generated with $u_{B,1}, u_{B,2} \sim U(-0.1, 0.1)$. 
Additionally, for each NO, the corresponding NOEM is applied to solve Eq.~\eqref{eq:ex2.1_setup} with 100 BCs randomly sampled from $u_1, u_2 \sim U(-0.1, 0.1)$. We observe a strong linear correlation ($R^2 = 0.789$) between NO prediction error and NOEM solution error (Fig.~\ref{fig:ex21_results}D). This result indicates that to have a good accuracy of the NOEM solution on the entire domain, we need to ensure a good NO accuracy on a small subdomain.

\subsubsection{Multiscale source term}
\label{sec:ex2.2}

\begin{figure}[htbp]
    \centering
    \includegraphics[width=0.9\linewidth]{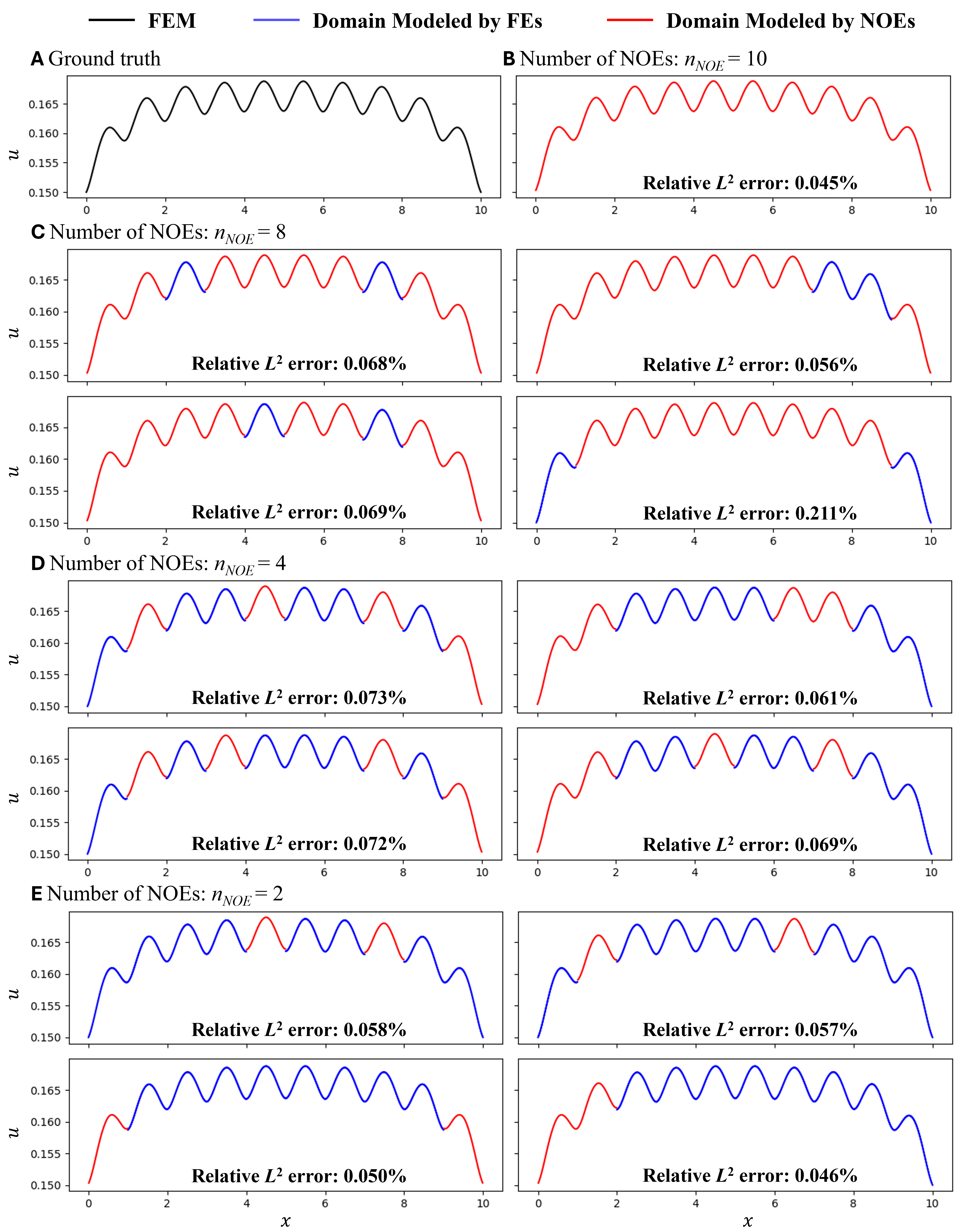}
    \caption{\textbf{Multiscale source-term problem (Section~\ref{sec:ex2.2}).} (\textbf{A}) FEM solution. (\textbf{B}) NOEM solution using 10 NOEs. (\textbf{C}) Examples of NOEM solutions using 8 NOEs. (\textbf{D}) Examples of NOEM solutions using 4 NOEs. (\textbf{E}) Examples of NOEM solutions using 2 NOEs. (C, D, and E) The locations of NOEs (red color) are randomly chosen. For instance, the example at the bottom left of (E) uses $[0,1]$ and $[9,10]$ by the NOEs, and the rest subdomains are modeled by the conventional FEs.}
    \label{fig:ex22_results}
\end{figure}

We also test another multiscale problem:
\begin{equation*}
\label{ex22_setup}
    \begin{aligned}
        -\frac{d^2u}{dx^2} + \frac{du}{dx} &= f(x), \quad && \text{for } x \in [0, 10], \\
        u(x) &= 0,  \quad && \text{at } x = 0, \\
        u(x) &= 0,  \quad && \text{at } x = 10. \\
    \end{aligned}
\end{equation*}
where $f(x) = (x\mod{1}) (1 - x\mod{1})$. For the implementation of the NOEM, the domain $[0, 10]$ is divided into 10 subdomains: $\{[i,i+1]\}_{i=0}^9$. Within each subdomain, one of the two modeling strategies is deployed: (1) 100 conventional FEs, or (2) one NOE. For NOE subdomains, a DeepONet is utilized to learn the solution operator mapping from the BC to the solution of the subdomain:
\begin{equation*}
\begin{aligned}
    \mathcal{G}: \boldsymbol{u}_{\text{BC}} \mapsto u_{\mathcal{G}},
\end{aligned}
\end{equation*}
where $\boldsymbol{u}_{\text{BC}} = \{u_{B,1},u_{B,2} \}$ represents the boundary condition.
We test NOEM with different numbers of NOEs, denoted as $n_{NOE}$, varying from 2 to 10. The solutions in Fig.~\ref{fig:ex22_results} show that the accuracy of the NOEM is robust to the choice of the locations of the subdomains modeled by the NOEs.

\subsection{Heat transfer on complex geometries with varying scales}
\label{sec:ex3}

Two-dimensional heat transfer problems defined on complex geometries $\Omega$ with varying scales are investigated here to validate the computational efficiency and model reusability of the NOEM. The governing equation of the heat transfer problem is formulated as
\begin{equation*}
-\nabla \cdot (K\nabla u(x,y)) = 0 \quad \text{for } (x,y) \in \Omega,
\end{equation*}
where $K=1$ is the thermal diffusivity.

\subsubsection{Rectangular domain with a single hole}
\begin{figure}[htbp]
    \centering
    \includegraphics[width=\linewidth]{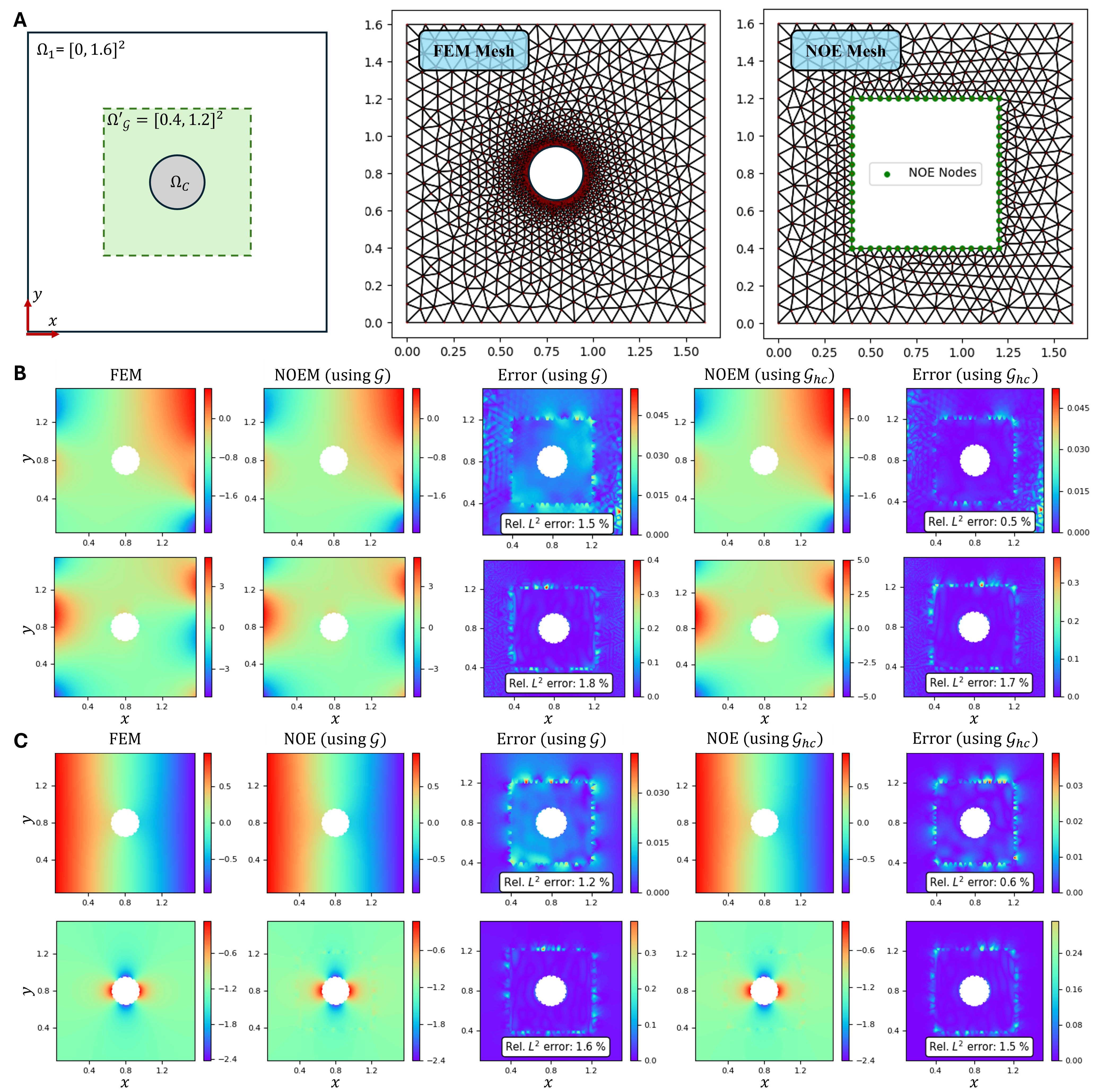}
    \caption{\textbf{Heat transfer on the rectangular domain with a single hole.} (\textbf{A}) In the computational domain, the gray area and the green areas denote the central circular opening $\Omega_C$, the domain modeled by the NOE$\Omega_{\mathcal{G}}$, respectively. The meshes used in the FEM and the NOEM are visualized in the middle and right figure panels. 
    (\textbf{B}) FEM and the NOEM solutions for Boundary Condition 1. The first and second rows are the PDE solutions $u$ and its derivatives $\frac{\partial u}{\partial x}$.
    (\textbf{C}) Boundary Condition 2.}
    \label{fig:ex3_1}
\end{figure}

We investigate the heat transfer problem in a rectangular domain with a central circular opening (Fig.~\ref{fig:ex3_1}A):
\begin{align*}
    \Omega = \Omega_1 \setminus \Omega_C,
\end{align*}
where $\Omega_1 = [0,1.6]^2$ is a rectangular area, and $\Omega_C = \{(x,y) \mid (x-0.8)^2 + (y-0.8)^2 \leq 0.15^2\}$ is the central circular area with a radius of 0.15.
Two different boundary conditions are investigated here. In the first case (Boundary Condition 1), the Dirichlet boundary conditions $v \sim \mathcal{GP}$ with a correlation length $l=0.3$ are applied to the left and right boundaries of the rectangular area, $\Gamma= \{x \in \{0, 1.6\} \text{ and } y \in [0, 1.6]\}$. For the remaining boundaries of $\Omega$, zero Neumann boundary conditions are employed. Thus, Boundary Condition 1 is
\begin{equation}
    \label{eq:ex3_bc}
    \begin{aligned}
        u(x, y) = v(x, y) \quad &\text{for } (x, y) \in \Gamma, \\
        \partial_n u(x, y) = 0 \quad &\text{for } (x, y) \in \partial \Omega \setminus \Gamma.
    \end{aligned}
\end{equation}
Another boundary condition, referred to as Boundary Condition 2, is
\begin{equation*}
    \label{eq:ex3_bc2}
    \begin{aligned}
        u(x, y) = 1,& \quad &&\text{for } x=0, \\
        u(x, y) = -1,& \quad &&\text{for } x=1.6, \\
        \partial_n u(x, y) = 0,& \quad &&\text{for } (x, y) \in \partial \Omega \setminus \Gamma.
    \end{aligned}
\end{equation*}

The meshes generated for the FEM and the NOEM are depicted in Fig.~\ref{fig:ex3_1}A. Two NOs, including a vanilla DeepONet ($\mathcal{G}$) and a hard-constraint DeepONet ($\mathcal{G}_{hc}$), are trained to map the Dirichlet boundary conditions to the solutions over $\Omega_\mathcal{G} = {\Omega'}_\mathcal{G} \setminus \Omega_C$ (the green area in Fig.~\ref{fig:ex3_1}A), where ${\Omega'}_\mathcal{G} = [0.4, 1.2]^2$ is a rectangular area.

Comparisons between the FEM and the NOEM for both the solution $u$ and its derivatives $\frac{\partial u}{\partial x}$ are visualized in Figs.~\ref{fig:ex3_1}B and C for the two BC cases. The NOEM demonstrates good accuracy in solving both PDE solutions and derivatives, This example affirms the NOEM's effectiveness in solving PDEs on complex geometries using a coarse mesh.

\subsubsection{Rectangular domain with multiple holes}
\label{sec:ex3.2}

\begin{figure}[htbp]
    \centering
    \includegraphics[width=0.95\linewidth]{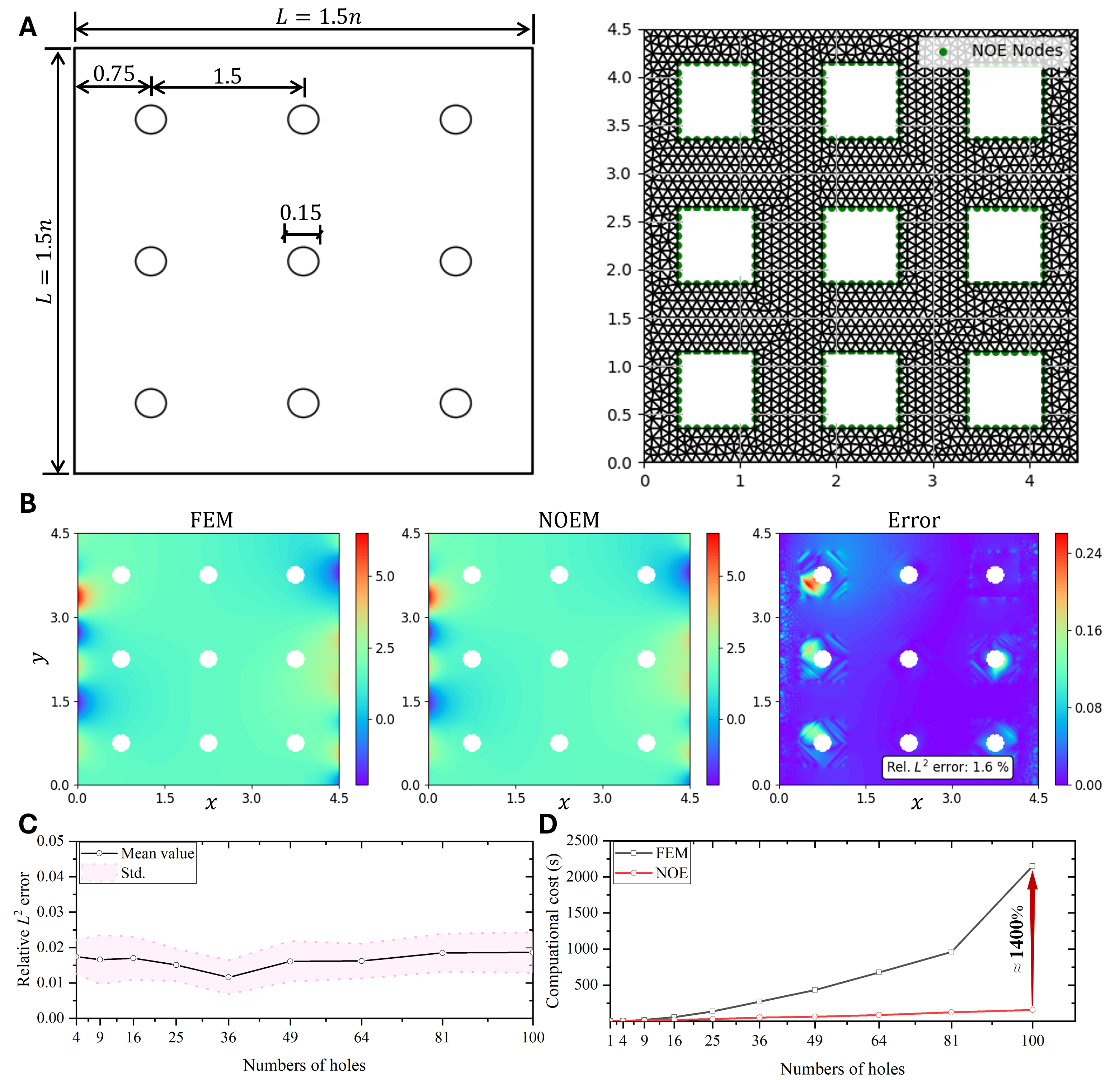}
    \caption{\textbf{Heat transfer on a rectangular domain with multiple holes.} (\textbf{A}) A rectangular domain containing $3\times3=9$ holes. The mesh for the NOEM, where the boxes outlined by the green dots are the subdomains modeled by the NOEs. (\textbf{B}) Comparison between FEM and NOEM. The relative $L^2$ error of the NOEM is 1.6\%. (\textbf{C}) Relative $L^2$ errors for different numbers of holes. (\textbf{D}) Comparison of computational time between FEM and NOEM for different numbers of holes.}
    \label{fig:ex3.2_results}
\end{figure}

In this example, we further investigate the NOEM for the heat transfer of a rectangular domain with multiple holes (Fig.~\ref{fig:ex3.2_results}A) to demonstrate the efficiency and scalability of the NOEM. When there are $n$ rows and columns of holes, the computational domain is $\Omega_n = [1.5n, 1.5n]^2$. The employed boundary conditions are similar with the one in Eq.~\eqref{eq:ex3_bc}, where the Dirichlet boundary condition is randomly generated and applied to the left and right boundaries, while the zero Neumann boundary condition is employed on the remaining parts of the boundary. 

When the rectangular domain contains $3\times3$ holes ($n = 3$), the mesh of the NOEM is visualized in Fig.~\ref{fig:ex3.2_results}A. It is important to note that we directly reuse the DeepONet trained in the previous single-hole example in the NOEM without additional tuning or training. The PDE solutions of the FEM and the NOEM are presented in Fig.~\ref{fig:ex3.2_results}B. The NOEs with the previously trained DeepONet demonstrate accurate results, underscoring the model reusability within the proposed NOEM.

Subsequent tests were conducted to assess the robustness and efficiency of the NOEM. We test different domains with different numbers of holes by choosing $n$ from 2 to 10. For each case, 20 boundary conditions were randomly generated and used to solve the PDE. The relative $L^2$ errors of the NOEM are summarized in Fig.~\ref{fig:ex3.2_results}C, illustrating stable performance across different $n$ values, which indicates its robustness and scalability for problems of varying scales. Additionally, the computational costs of the FEM and the NOEM are presented in Fig.~\ref{fig:ex3.2_results}D. NOEM is significantly faster than FEM, especially for large-scale problems. We observe around 14 times speedup in the computational cost for 100 holes, indicating the efficiency and scalability of the NOEM.

\subsection{Darcy flow with complex permeability fields}
\label{sec:ex4}
In this example, we consider a Darcy flow problem on a rectangular domain:
\begin{equation*}
\begin{aligned}
    -\nabla \cdot (K(x,y) \nabla u(x,y)) = 0& \quad \text{for } (x,y) \in \Omega_1  \\
    u(x,y) = v(x, y)& \quad \text{for } x = 0 \text{ and } x=2 \\
    K(x,y) \partial_n u(x,y) = 0& \quad \text{for } y = 0 \text{ and } y = 2 
\end{aligned}
\end{equation*}
where $\Omega_1 = [0,2]^2$ is a rectangular area; $u$ denotes the pressure; $v$ is a function in the Dirichlet boundary conditions. $K(x,y)$ is the permeability field defined as
\[
K(x,y) = 
\begin{cases} 
1 & \text{for } (x,y) \in \Omega_1 \setminus \Omega_2, \\
K'(x,y) & \text{for } x \in \Omega_2,
\end{cases}
\]
with \(\Omega_2 = [0.5,1.5]^2\). Four types of permeability fields are investigated as follows.

\begin{figure}[htbp]
    \centering
    \includegraphics[width=\linewidth]{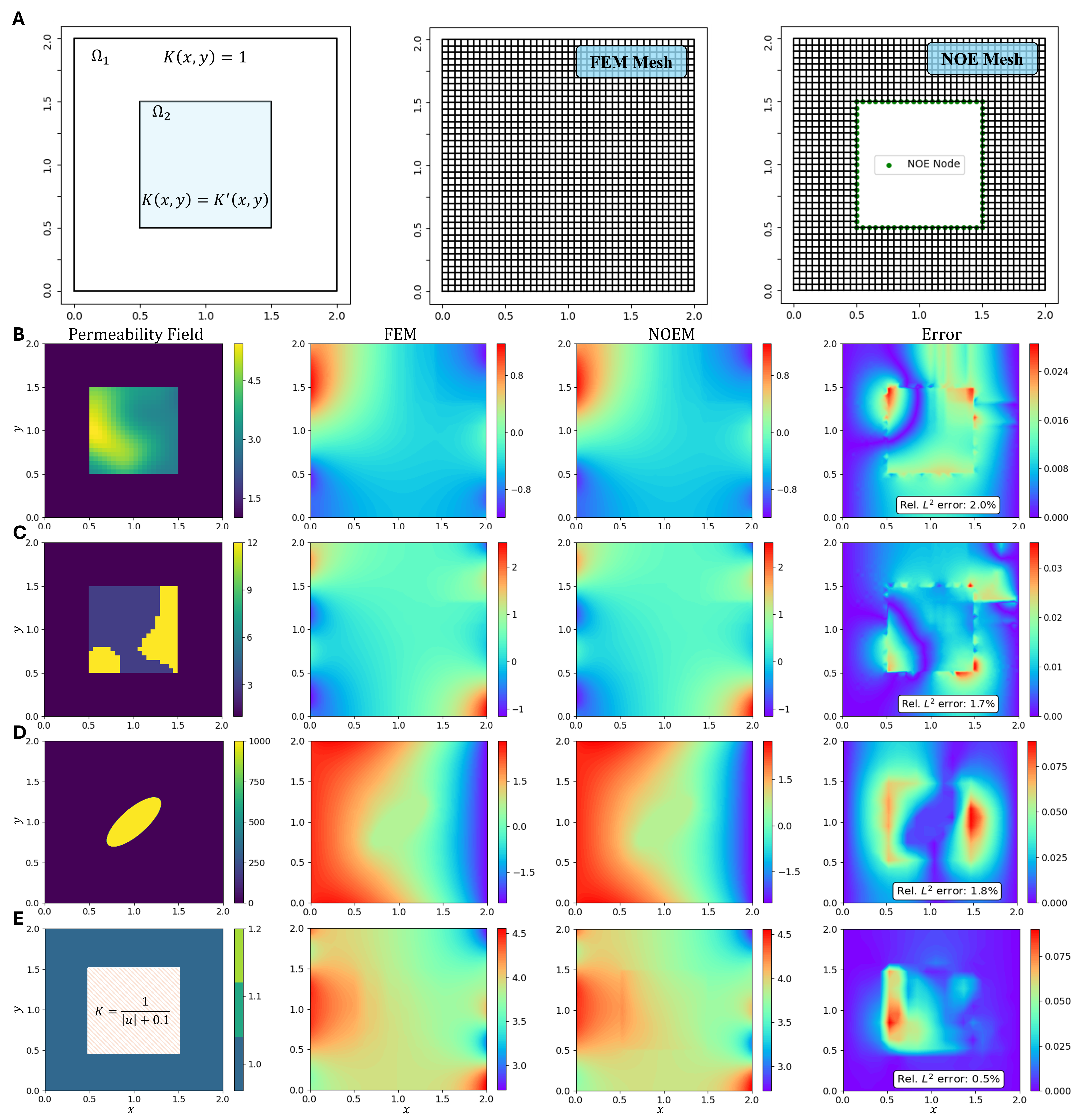}
    \caption{\textbf{Darcy flow with different complex permeability fields.} (\textbf{A}) The computational domain, and the meshes for the FEM and the NOEM. (\textbf{B}) An example of a continuous permeability field. (\textbf{C}) An example of a piecewise constant permeability field. (\textbf{D}) An example of a permeability field parameterized by an elliptic shape with $a=0.4,b=0.15,\text{and } \theta=\frac{\pi}{4}$. (\textbf{E}) An example of a nonlinear permeability field.}
    \label{fig:ex4_results}
\end{figure}

The FEM with a $40 \times 40$ mesh is utilized to obtain the reference solution. In the NOEM, the entire subdomain $\Omega_2$ is modeled using a single NOE. The meshes for the FEM and NOEMs are depicted in Fig.~\ref{fig:ex4_results}A.

\subsubsection{Continuous permeability field}

We first consider continuous permeability fields:
\begin{equation*}
    K'(x,y) = \exp(w(x,y)), \quad w(x,y) \sim \mathcal{GP} ( 0, k_l((x,y),(x',y')) ),
\end{equation*}
where $\mathcal{GP}(0, k_l((x,y),(x',y')))$ is a GP with a correlation length $l=0.3$. For the NOEM, a MIONet is employed to learn the mapping from the combined inputs of boundary conditions and the permeability field to the pressure field:
\begin{equation}
    \label{eq:ex4_mionet}
    \mathcal{G}: (u|_{\partial{\Omega}_\mathcal{G}}, K') \mapsto u_{\mathcal{G}}
\end{equation}
One example of a random permeability field and the PDE solutions from the FEM and the NOEM are shown in Fig.~\ref{fig:ex4_results}B. The relative $L^2$ error of the NOEM is approximately 2\%.

\subsubsection{Piecewise constant permeability field}

Then, we consider piecewise constant permeability fields:
\begin{equation*}
\begin{aligned}
    K'(x,y) = m(w(x,y)), \\
m(v) = 
\begin{cases} 
3 & \text{for } w \leq 0, \\
12 & \text{for } w > 0.
\end{cases}
\end{aligned}
\end{equation*}
Similar to Eq.~\eqref{eq:ex4_mionet}, we also use MIONet to learn $ \mathcal{G}: (u|_{\partial{\Omega}_\mathcal{G}}, K') \mapsto u_{\mathcal{G}}$, while the permeability field $K'$ is piecewise constants. One example of a random permeability field and the solutions of the FEM and the NOEM are presented in Fig.~\ref{fig:ex4_results}C, where the relative $L^2$ error of NOEM is smaller than 2\%.

\subsubsection{Permeability field parameterized by an elliptic shape}

Next, we test permeability fields parameterized by an elliptic shape 
\begin{equation*}
    K'(x,y) = 
    \begin{cases} 
        1 & \text{for } (x,y) \notin \Omega_E, \\
        1000 & \text{for } (x,y) \in \Omega_E,
    \end{cases}
\end{equation*}
where
\begin{equation*}
    \Omega_E = \left\{(x,y) \mid 
    \left[\frac{(x-1)\cos\theta + (y-1)\sin\theta}{a}\right]^2 + 
    \left[\frac{-(x-1)\sin\theta + (y-1)\cos\theta}{b}\right]^2 \leq 1\right\}
\end{equation*}
is an elliptic domain characterized by $a$, $b$, and $\theta$. The Dirichlet boundary conditions at the left and right boundaries are selected as $v(0, y) = 2.5$ and $v(2, y) = -2.5$.

A MIONet is trained to approximate the operator $\mathcal{G}$ mapping the boundary condition $u|_{\partial{\Omega}_\mathcal{G}}$ and the elliptic shape parameter $\boldsymbol{\theta}=\{a,b,\theta\}$ to the pressure field:
\begin{equation*}
    \mathcal{G}: (u|_{\partial{\Omega}_\mathcal{G}}, \boldsymbol{\theta}) \mapsto u_{\mathcal{G}}.
\end{equation*}

When $a=0.4,b=0.15,\theta=\frac{\pi}{4}$, the comparison between the FEM and the NOEM is provided in Fig.~\ref{fig:ex4_results}D, which shows the relative $L^2$ error of the NOEM is around 1.8\%.

\subsubsection{Nonlinear permeability field}
\label{sec:ex4.4}

We further test the Darcy flow problem with a nonlinear permeability field
\begin{equation*}
    K'(x,y) = \frac{1}{|u(x,y)| + 0.1}.
\end{equation*}

A DeepONet is trained to approximate the mapping between the boundary condition to the PDE  solution:
\begin{equation*}
    \mathcal{G}: u|_{\partial{\Omega}_\mathcal{G}} \mapsto u_{\mathcal{G}}.
\end{equation*}

For the training dataset, the boundary condition $u|_{\partial{\Omega}_\mathcal{G}}$ is generated from a GP, whose mean value and correlation length are 5 and 0.3, respectively. 
The FEM results and the NOEM results are shown in Fig.~\ref{fig:ex4_results}E. The relative $L^2$ error of the NOEM is around 0.5\%, indicating the robustness of the applicability in solving nonlinear PDE problems.
\section{Conclusions}
\label{sec:conclusions}

In this study, we introduce the neural-operator element method (NOEM), a hybrid numerical framework that synergistically combines operator learning with the finite element method (FEM) to leverage the strengths of both approaches. The NOEM employs a pretrained neural operator (NO) to construct NO elements (NOEs) that effectively model subdomains characterized by complex properties or geometries, allowing for accurate simulations using coarser meshes without compromising precision. NOEs significantly reduce the complexity of the problem, making NOEM applicable for large-scale problems.

This work offers a bridge between the FEM and the emerging operator learning research to benefit them from each other. In the future, more studies can be conducted to further improve NOEM. We can extend the NOEM with physics-informed machine learning, thereby enhancing overall performance and reducing the requirement of training data. We will apply the NOEM to address time-dependent PDE problems, thus broadening its applicability. We may also utilize emerging foundation model techniques for operator learning \cite{liu2023prose, hao2023gnot} to develop versatile NOEs capable of modeling diverse systems and scenarios.

\section{Acknowledgments}
The work described in this paper was partially supported by grants from the Research Grants Council of the Hong Kong Special Administrative Region through the projects "INTACT: Intelligent tropical-storm-resilient system for coastal cities (T22-501/23-R)".

\appendix
\section{Auxiliary functions for constructing neural operators with hard-constraint boundary conditions}
\label{sec:app_hcno}

Here, we present the method to implement hard-constrained BCs for NOs. Applying hard-constraint BC is challenging for PDEs defined on complex two-dimensional or three-dimensional domains, because it is difficult to analytically formulate auxiliary functions in Eq.~\eqref{eq:hcno}. To address this, various efforts have been made to construct these auxiliary functions. For example, Sheng and Yang \cite{sheng2021spline_hc} utilized spline interpolation to construct the function $\alpha$ for second-order boundary-value problems on intricate geometries. Sukumar and Srivastava \cite{sukumar2022exact} proposed a method to construct $\alpha$ using R-functions based on distance fields, which further ensures the differentiability of the auxiliary function.

Although these approaches have shown promising results~\cite{yazdani2020systems,lu2021hcpinn}, they are predominantly developed for physics-informed neural networks, rather than operator learning. Consequently, these studies focus primarily on constructing the function $\alpha$, and taking $\beta_g = g$ as the boundary condition, under the assumption that $g$ can be easily represented. However, when considering an NO with varying boundary conditions, such as for learning $\mathcal{G}: g \mapsto u$, where $g$ and $u$ represent the Dirichlet boundary condition and the corresponding PDE solution, respectively, the formulation of $\beta_g$ might not be straightforward. In our NOEM, the boundary conditions $g$ are often parameterized by the sensor values $\bm{g}:=\{g(\xi_i) \}_{i=1}^m$ at sensor locations $\Gamma:=\{\xi_i \in \partial\Omega\}_{i=1}^m$. In these cases, formulating the auxiliary function $\beta_g$ is challenging, as $g$ is not analytically given but only expressed through discretized sensor values.

We develop a novel method to construct $\beta_g$ using piecewise polynomials as $\beta_g = p_g$. Let us consider a PDE defined on $\Omega$ with the boundary condition given in Eq.~\eqref{eq:hc_pde_setting}. To enforce the NO compliant with Eq.~\eqref{eq:hc_pde_setting}, the auxiliary function in Eq.~\eqref{eq:hcno} should satisfy
\begin{equation} 
\label{eq:hc_aux_func_req}
    \beta_g(\xi_i) = g(\xi_i) \quad \text{for } \xi_i \in \Gamma.
\end{equation}
We assume that the domain $\Omega$ is discretized into a mesh $\mathcal{H}$. On this mesh, we define a piecewise polynomial, $\mathcal{P}_{\mathcal{H}}(\cdot; \bm{N}_{\partial \Omega}, \bm{N}_\Omega)$, which is constructed based on the nodal values at the mesh nodes. Specifically, $\bm{N}_{\partial \Omega}$ contains the values of the nodes located on the boundary $\gamma^{\partial\Omega}:=\left\{ \xi_i^{\partial\Omega} \in \partial\Omega \right\}_{i=i}^m$, and $\bm{N}_\Omega$ includes the values at all other nodes situated strictly within the domain, denoted by $\gamma^{\Omega}:=\left\{ \xi_i^\Omega \in \Omega \right\}_{i=i}^M$. To satisfy Eq.~\eqref{eq:hc_aux_func_req}, we first align the nodes in $\mathcal{H}$ on the boundary with the sensor locations parametrizing the boundary condition as $\gamma^{\partial\Omega}=\Gamma$. Then, we can construct $p_g$ as
\begin{equation*}
    p_g(\cdot) = \mathcal{P}_{\mathcal{H}}(\cdot; \bm{N}_{\partial \Omega} = \bm{g},\bm{N}_\Omega = \mathcal{N}(\bm{g})),
\end{equation*}
where $\mathcal{N}: \mathbb{R}^m \rightarrow \mathbb{R}^M$ is an mapping from $\bm{g}$ to the nodal values on $\gamma^{\Omega}$.

For simplicity, one could set all nodal values within the domain, $\left\{ \xi_i^\Omega \right\}_{i=1}^M$, to zero as $\mathcal{N}(\bm{g}) = \{0\}_{i=i}^M$. However, we adopt a different approach to construct $\mathcal{N}$. First, we train a NO without the hard-constrained BC to map from the boundary condition to the solution as $\mathcal{G}_{\text{NO}}': g \mapsto u$. Subsequently, we define $\mathcal{N}$ as
\begin{equation*}
    \mathcal{N}(\bm{g}) = \{\mathcal{G}_{\text{NO}}'[\bm{g}](\xi_i^\Omega) | \xi_i^{\Omega} \in \gamma^{\Omega}\}_{i=1}^M.
\end{equation*}

In NOEM, to ensure the continuity of the NOEM solution, we employ the same types of piecewise polynomials for $\beta_g$ and the FEs in the NOEM. Furthermore, on the interface between the domains modeled by the conventional FEs and the NOEs, we align the sensor locations of the hard-constraint NO with the FE node. This alignment guarantees the continuity of the NOEM solution across the interface.

\section{Learning neural operators}
\label{sec:no_training}
This section provides details for training NOs that construct NOEs.
Consider the following PDE:
\begin{equation*} \label{noe_pde_solution}
    \begin{aligned}
        \mathcal{D}[u|_{\Omega^\text{NOE}_j}] = 0, \quad& \text{for } x \in \Omega^\text{NOE}_j, \\
        u|_{\Omega^\text{NOE}_j} = u|_{\partial\Omega^\text{NOE}_j}, \quad& \text{for } x \in \partial \Omega^\text{NOE}_j,
    \end{aligned}
\end{equation*}
where $\mathcal{D}$ is the underlying differential operator of the PDE. A NO is trained to approximate the operator $\mathcal{G}$ mapping from the boundary conditions to the corresponding solution of the PDE as
\begin{align*}
    \mathcal{G}:u|_{\partial\Omega^\text{NOE}_j} \mapsto u|_{\Omega^\text{NOE}_j}.
\end{align*}

To train this NO, we need to generate a training dataset $\{u_{\partial \Omega^\text{NOE}_j}^{(i)}, u_{\Omega^\text{NOE}_j}^{(i)}\}_{i=1}^{N}$. We first sample many BCs $u|_{\partial\Omega^\text{NOE}_j}$ from the function space of interest, and then obtain the corresponding PDE solutions by a conventional numerical solver, e.g., the FEM.
Then, the NO with trainable parameters $\theta$ is trained to fit the dataset. A popular loss function is the mean squared error (MSE).

For the NOs utilized in our numerical examples, the subnetworks employed can be classified into two types: (1) MLP, which comprise three to four hidden layers, each containing 50 to 200 neurons; and (2) CNN models, consisting of two convolutional layers with a kernel size of 5 and a stride of 2, followed by two fully-connected layers. 
We employed the Adam optimizer with a learning rate of 0.001 for the model training. 
During model training, we set the maximum training epoch as 1,000,000 and employed an early-stop criterion that terminates training if there is no improvement in model performance within a span of 10,000 epochs. 
The training dataset sizes are 1,000 and 5,000 for the NOs employed for 1D and 2D problems, respectively.

\section{Implementation details of NOEM for solving PDEs}
\label{sec:noe_imp}

We present the process of using NOEM with a pre-trained NO $\mathcal{G}$ to solve a PDE. We consider the functional $J$ is written as
\begin{equation*}
    J[u]=\int_{\Omega} F[u](x) dx,
    \label{eq:noe2}
\end{equation*}
where $F$ is an appropriate operator associated with the PDE. Using the representation in Eq.~\eqref{eqn:solution}, the energy functional can be written as
\begin{equation*}
    J[q(\cdot;\bm{c})] = 
    \sum_{i \in I_\text{FE}}
    \int_{\Omega^\text{FE}_i} F[\phi_i^\text{FE}(\cdot;\bm{\alpha}_i)](x) dx
    + \sum_{j \in I_\text{NOE}}
    \int_{\Omega^\text{NOE}_j} F[\phi_j^\text{NOE}(\cdot;\bm{\beta}_j)](x) dx.
\end{equation*}
We then apply numerical quadrature rules to approximate the integral terms for NOEs:
\begin{equation*} \label{def:actual-loss}
    J[q(\cdot;\bm{c})]
    \approx \sum_{i \in I_\text{FE}}
    \int_{\Omega^\text{FE}_i} F[\phi_i^\text{FE}(\cdot;\bm{\alpha}_i)](x) dx
    + \sum_{j \in I_\text{NOE}}
    \left[\sum_{s} w_s^{(j)} F[\phi_j^\text{NOE}(\cdot;\bm{\beta}_j)](z_s^{(j)})\right],
\end{equation*}
where $\{z_s^{(j)},w_s^{(j)}\}$ is a quadrature rule for the domain $\Omega^\text{NOE}_j$.

Newton's method is then employed in this study to solve Eq.~\eqref{eq:noe3}. Newton's method relies on the Hessian matrix and gradient vector of $J[q(\cdot;\bm{c})]$, which is assembled by the Hessian matrix and gradient vector of the energy norm of each element as
\begin{align*}
    J[\phi_i^\text{FE}(\cdot;\bm{\alpha}_i)] = \int_{\Omega^\text{FE}_i} F[\phi_i^\text{FE}(\cdot;\bm{\alpha}_i)](x) dx,
    &\quad
    J[\phi_j^\text{NOE}(\cdot;\bm{\beta}_j)] =  \left[\sum_{s} w_s^{(j)} F[\phi_j^\text{NOE}(\cdot;\bm{\beta}_j)](z_s^{(j)})\right], \\
    \boldsymbol{g}_i^\text{FE} \left(\bm{\alpha}_i \right) = \frac{\partial J[\phi_i^\text{FE}(\cdot;\bm{\alpha}_i)]}{\partial \bm{\alpha}_i}, 
    &\quad 
    \boldsymbol{k}_i^\text{FE} \left(\bm{\alpha}_i \right) = \frac{\partial \boldsymbol{g}_i^\text{FE} \left(\bm{\alpha}_i \right)}{\partial \bm{\alpha}_i}, \\
    \boldsymbol{g}_j^\text{NOE} \left(\bm{\beta}_j \right) = \frac{\partial J[\phi_j^\text{NOE}(\cdot;\bm{\beta}_j)]}{\partial \bm{\beta}_j}, 
    &\quad 
    \boldsymbol{k}_j^\text{NOE} \left(\bm{\beta}_j \right) = \frac{\partial \boldsymbol{g}_j^\text{NOE}}{\partial \bm{\beta}_j}, 
\end{align*}
where $\boldsymbol{g}$ and $\boldsymbol{k}$ represent the gradient vector and Hessian matrix of elements $\phi_i^\text{FE}$ and $\phi_j^\text{NOE}$, respectively.
Since $\phi_i^{\text{FE}}$ are commonly expressed explicitly, the gradient vector $\boldsymbol{g}_i^{\text{FE}}$ and the Hessian matrix $\boldsymbol{k}_i^{\text{FE}}$ can be derived in closed form. For the NO, automatic differentiation (AD) is utilized to compute $\boldsymbol{g}_j^{\text{NOE}}$ and $\boldsymbol{k}_j^{\text{NOE}}$.
Subsequently, the overall gradient vector $\boldsymbol{G}(\bm{c})$ and Hessian matrix $\boldsymbol{K}(\bm{c})$ of $J[q(\cdot; \bm{c})]$ are assembled from these individual components. The assembled $\boldsymbol{G}(\bm{c})$ and $\boldsymbol{K}(\bm{c})$ is employed in Newton's method to solve Eq.~\eqref{eq:noe3}, as detailed in Algorithm~\ref{alg:NOEM}.

\begin{algorithm}[htbp]
\caption{\textbf{NOEM for solving a PDE.}}
\label{alg:NOEM}
\begin{algorithmic}
\State \textbf{Meshing:} Discretize the computational domain into $\Omega^\text{FE} = \{\Omega^\text{FE}_{i}\}_{i\in I_\text{FE}}$ and $\Omega^\text{NOE}= \{\Omega^\text{NOE}_{i}\}_{i\in I_\text{NOE}}$.
\State \textbf{Initialization:} Initialize nodal value vector, $\bm{c}$, according to boundary conditions.
\State \textbf{Main Loop:}
\For{$iter = 0$ to $iter_{\text{max}}$}
    \State Compute $\boldsymbol{k}_i^\text{FE}$ for each conventional FE.
    \State Compute $\boldsymbol{k}_j^\text{NOE}$ using the AD for each NOE.
    \State Assemble the Hessian matrix, $\boldsymbol{K}(\bm{c})$, according to the boundary conditions.
    \State Compute $\boldsymbol{g}_i^\text{FE}$ for each conventional FE.
    \State Compute $\boldsymbol{g}_j^\text{NOE}$ using the AD for each NOE.
    \State Assemble the gradient vector, $\boldsymbol{G}(\bm{c})$, according to the boundary conditions.
    \State Solve for $\Delta \bm{c} = \left( \boldsymbol{K}(\bm{c}) \right)^{-1} \boldsymbol{G}(\bm{c})$.
    \State Update the nodal value vector as $\bm{c} \leftarrow \Delta \bm{c} + \bm{c}$.
    \State If $\frac{\| \Delta \bm{c} \|_2}{\| \bm{c} \|_2} \leq \text{Tolerance}$, \textbf{stop}.
\EndFor
\end{algorithmic}
\end{algorithm}

Although the NO inputs in our study use sensor values at the NOE mesh grid points (illustrated by the green dots in Fig.~\ref{fig:method_ill}A), other discretization methods could also be utilized. When employing another representation $\bm{\beta}'_j$, the gradient vector $\boldsymbol{g}_j^\text{NOE}$ and Hessian matrix $\boldsymbol{k}_j^\text{NOE}$ with respect to the values at the NOE mesh nodes $\bm{\beta}_j$ can be computed via leveraging the Jacobian matrix of $\bm{\beta}'_j$ with respect to $\bm{\beta}_j$.

\bibliographystyle{unsrt}
\bibliography{main}

\end{document}